\documentclass[prb,amsmath,amssymb,shownopacs,twocolumn]{revtex4}
\usepackage{graphicx}

\newcommand{\beq}{\begin{equation}}
\newcommand{\eneq}{\end{equation}}
\newcommand{\bea}{\begin{eqnarray}}
\newcommand{\eea}{\end{eqnarray}}

\input{epsf}

\begin{document}



\title{Reply to the comment to Phy. Rev. Lett. 86, 3392 (2001) 
(cond-mat/0511607)}





\maketitle














We provide our reply to the comment by Greiter and Schuricht (GS) \cite{gs0}.
Let us first stress that there is no doubt about the mathematical
correctness of our derivation.  We now show that our interpretation
of the physics of spinons is correct. Against the existence of
spinon attraction, GS recall that spinons constitute an ``ideal gas
of half fermions'' \cite{hald}. In reality, spinons feel a
statistical interaction associated to a rule for progressively
filling single-particle states. The fully dressed $S$-matrix for
spinons \cite{essler} in the Haldane-Shastry model (HSM) \cite{hs}
takes the trivial form $S = \pm i$ times the identity ${\bf I}$.
(This is well known - GS strangely fail to quote the result we
derived on page 9 of the second paper of Ref.\cite{us} which is
equivalent to computing the $S$-matrix associated to exchange
statistics of semions - which can also be done by Asymptotic Bethe
Ansatz - there is an extra $i$ in our formula: it comes from a
misprint in the final version of the paper). Triviality of the
$S$-matrix means that spinons are alleged asymptotic states of the
HSM. Their long-distance behavior is not affected by their dynamical
short range attraction. In our paper, we explicitly show the
short-distance effects of the interaction on two
spinon-wavefunctions \cite{us}. We now consider the points raised by
GS separately.

1. GS claim our interpretation of the $p_{mn} ( e^{ i \theta } )$ as
two-spinon relative wavefunction is ambiguous, because the
$p_{mn}$'s are defined by expanding the (overcomplete) set of states
$\Psi_{\alpha \beta}$ in the (basis of) energy eigenfunctions
$\Phi_{mn}$. This statement is false. In real space, spinons are
nonlocal excitations, with typical size of the order of the lattice
step  \cite{us}. The $\Psi_{\alpha \beta}$'s should be thought of as
the lattice version of two-particle coherent states. The
overcompleteness of the $\Psi_{\alpha \beta}$'s is, therefore, the
lattice version of the usual overcompleteness of coherent states.
Two spinons at the same site $\xi$ correspond to a localized spin-1
excitation, unambiguously described by $\Psi_\xi ( z_1 , \ldots ,
z_{N/2 - 1} )$ \cite{us}. The physical meaning of $p_{mn} ( 1 )$ as
the probability enhancement when the two spinons
 are at the same site (and form the spin-1 excitation) is hence absolutely
  non-ambiguous, at odds to GS's
statement that the $| p_{mn} |^2$'s ``cannot be interpreted as
probability distribution''. We also find a similar short-range
attraction between a spinon and a holon  (in the supersymmetric
$t-J$ model with $1 / r^2$-interaction) although the states of a
localized spinon and a localized holon do not form an overcomplete
set \cite{us}. We wish to remark that using overcomplete sets of
states to treat nonlocal excitations as quantum-mechanical particles
is in fact widely used for Laughlin quasiholes, a fact which Greiter
correctly and explicitly points out in a prior publication
\cite{greit}. The spinon situation is no different.

2. GS assert that ``The second argument of BGL is that the last term
in their expression for the energy of the two-spinon states
represents ``a negative interaction contribution that becomes
negligibly small in the thermodynamic limit''. We clearly state that
the additional negative contribution to the two spinon energy is
negligibly small in the thermodynamic limit for most values of the
moments. The interaction, however, is not at all negligible for
spinons of large momenta difference $m-n ~ M$. The probability
enhancement $p_{mn}$ depends this difference. As we clearly show in
our plot, in the thermodynamic limit the spinon attraction become
most important.

3. The spinon-interaction is in fact evident from the real space
effective Hamiltonian for the two-spinon subspace which we derived
in equation[20] of \cite{us}. Besides the kinetic terms appropriate
for each of the spinons, the effective Hamiltonian also contains a
term proportional to the difference in the spinon velocities and
inversely proportional to the distance between spinons. This is
extremely suggestive of the gauge-like statistical forces between
fractionalized particles.

4. GS remark that ``it is not $ p_{mn} ( e^{ i \theta } ) $ as a
function of $\theta$ for fixed $m$ and $n$ which enter their
derivation of the DSS, but $p_{mn} (1)$ as a function of $m$ and
$n$''. This is exactly what we explicitly say in our paper. If
spinons were noninteracting, $p_{mn} (1)$ would not depend at all on
$m$ and $n$, and the distribution would be flat. On the contrary,
the strong dependence of $p_{mn} (1)$ on $m$ and $n$, which is a
consequence of the strong nonuniformity of $p_{mn} ( e^{ i \theta }
)$ on $\theta$, is the key ingredient, to obtain the square-root
singular threshold in the dynamical spin susceptibility. With the
experimental observation of the square-root singularity in ${\rm
KCuF}_3$, Tennant {\it et.al.} have observed fractional quantization
in spin chains  {\bf by looking at the main effect of the statical
interaction arising from quantum number fractionalization}.

\vspace{0.2cm}

B. A. Bernevig, D. Giuliano and R. B. Laughlin

\end{document}